**The Non-linear Osmotic Brush Regime: Experiments, Simulations and Scaling Theory**

*Heiko Ahrens[1], Stephan Förster[2], Christiane A. Helm[1], N. Arun Kumar[4], Ali Naji[3], Roland R. Netz[3], Christian Seidel[4]*

[1] Institut für Physik, Ernst-Moritz-Arndt Universität, F.-L.-Jahn-Str. 16, 17487 Greifswald, Germany
[2] Institut für Physikalische Chemie, Universität Hamburg, Bundesstraße 45, 20146 Hamburg, Germany
[3] Sektion Physik, Ludwig-Maximilians University, Theresienstr. 37, 80333 Munich, Germany
[4] Max-Planck-Institut für Kolloid- und Grenzflächenforschung, Am Mühlenberg 1, 14476 Potsdam, Germany

*Abstract*
We experimentally and theoretically consider highly condensed planar brushes made of charged polymers. Using x-ray reflectivity on polyelectrolytes which are anchored at the water-air interface, it is shown that such strongly stretched brushes show a slight but detectable height variation upon lateral compression. This stands in contrast to the well-accepted scaling relation in the so-called osmotic brush regime, which predicts the brush height to be independent of the grafting density. Similar effects are seen in simulations on highly compressed charged brushes. Scaling arguments which go beyond the linear approximation for the entropy of confined counterions and for weak chain-stretching are able to explain those findings on a semi-quantitative level.

## 1. Introduction

Polyelectrolytes anchored on surfaces are important for their many applications and also form a challenging topic for pure science[1-23]. Since those charged brushes typically trap their counterions and form a layer of very high internal salt concentration, their structure and behavior is rather insensitive to the amount of externally added salt. This gives rise to a wide range of applications for stabilization and surface functionalization of charged and neutral colloids[24,25]. One distinguishes a few scaling regimes. For not too high external salt concentrations and for fully charged polymer chains, the brush height results from a balance of the entropy of counterion confinement (which tends to swell the chain) and chain elasticity (which tries to shrink the chain). In the simplest scaling description of this so-called osmotic brush regime, which is the main topic of the present paper, the brush height is independent of the grafting density of chains.[13] In the accepted theory[13,15,16], the chain swelling is assumed to be weak. However, experiments[1-11] suggest chain swelling up to 70% of the contour length. In this paper we present experimental evidence showing that the brush height in fact increases slowly as the grafting density goes up. This is in agreement with Molecular Dynamics simulations[21] and can also be rationalized in terms of simple scaling arguments that take into account the finite excluded volume of polymer chains and the non-linear elasticity of strongly stretched polymer chains.[18] A weak increase of the osmotic brush height with grafting density has also been observed in recent experiments by Romet-Lemonne et al. that shows agreement with the non-linear scaling predictions.[11]

## 2. Experimental Section

**Experimental Materials and Methods**

Monolayers of the diblock copolymers poly-(ethyl ethylene)$_{144}$poly(styrene sulfonic acid)$_{136}$ (PEE$_{144}$PSS$_{136}$, degree of sulfonization 0.9) and poly-(ethyl ethylene)$_{114}$poly(styrene sulfonic acid)$_{83}$ (PEE$_{114}$PSS$_{83}$, degree of sulfonization 0.85) were investigated at the air/water-interface. The PEE is fluid at room temperature, therefore the PEE-PSS joints can attain a new equilibrium distribution after a change in the molecular area.

The x-ray set-up is home-built ($\lambda$=1.54Å),[26] its angular divergence is 0.012$^{\circ}$. For X-rays, the index of refraction of a certain material, $n$, depends linearly on the electron density $\rho$ and known constants (Thompson radius $r_o$), $n = 1 - r_0 \rho \lambda^2 / 2\pi$, and deviates for all materials at most by $\sim 10^{-5}$ from 1. The measured reflectivity $R$ can be understood as the Fresnel reflectivity $R_F$ of an infinitely sharp interface modulated by interference effects from the thin surface layer.[27] If the angle of incidence exceeds about two times the critical angle of total internal reflection, the reflectivity can be described by a kinematic approximation,

$$\frac{R}{R_F} = \left| \frac{1}{\rho_{sub}} \int \rho'(z) e^{iQ_z z} dz \right|^2 \qquad (1)$$

where $\rho_{sub}$ is the electron density of the bulk phase, $\rho'(z)$ the gradient of the electron density along the surface normal, and $Q_z$ the wave vector transfer normal the surface. Yet, to quantify the density profile, the exact formalism of optical transfer matrices between layers of different indices of refraction $n$ is used.[28] The surface layer is parameterized as consisting of different slabs (each with an electron density, a thickness, and a roughness parameter, which smoothes the transition to the next slab). In case of ambiguity, we chose the least structured profile (maximum entropy approach [29]), with which the measured reflectivity could be fitted.

**Results**

We will start with the data obtained from the molecules with the longer PSS block, because the reflectivity data are more structured. The PEE$_{144}$PSS$_{136}$ monolayer was prepared on a subphase containing 1 mM CsCl, which is a quite low concentration and more than two orders of magnitude smaller than the internal brush counterion concentration. X-ray reflectivity measurements were taken along the isotherm (cf. Fig.1). The reflectivity curves show a very clear structure. At low $Q_z$, a thick layer (the PSS block) causes up to three narrow maxima, separated by shallow minima. Upon monolayer compression, the contrast improves and more maxima can be distinguished. The position of the extrema is almost constant, there is no obvious indication of a change in the thickness of the PSS brush. A thinner layer (the PEE block) causes the minimum at high $Q_z$, which shifts strongly on compression to lower $Q_z$, indicating thickening of the PEE block.

In the exact matrix formalism, five different slabs are necessary: one for the PEE block, four for the PSS block. The PSS$_{136}$-brush causes three interference maxima, which are not equidistant. Therefore, three slabs are necessary to describe the decay of the segment density profile from the grafting interface towards the subphase. As expected for lateral pressures above $\pi_1$ [1,2], the hydrophobic PEE block behaves as an ultrathin melt exhibiting a constant electron density consistent with the mass density known from the bulk phase, and a linear relationship between grafting density and layer thickness. Directly beneath the hydrophobic layer is a ~12Å thick layer of high electron density, which is attributed to a flat monolayer of PSS-chains, which are adsorbed to the hydrophobic PEE block[1,2]. Above $\pi_1$, the lateral density of this monolayer is constant (1 monomer per 44 Å$^2$) and independent of the grafting density or the PSS length, the concentration or the type of cation. Only below $\pi_1$, the lateral density of the flatly adsorbed monolayer of PSS decreases.

To quantify the water and counterion concentration within the $PSS_{136}$ brush, we have to calculate the number of water molecules, $n_{H_2O}$, and $Cs^+$ ions, $n_{Cs}$, per polymer molecule:

$$E = A \sum_i \rho_i l_i = n_{SS} E_{SS} + n_{Cs} E_{Cs} + n_{H_2O} E_{H_2O}$$

$$V = A \sum_i l_i = n_{SS} V_{SS} + n_{Cs} V_{Cs} + n_{H_2O} V_{H_2O}$$ 

$\qquad\qquad\qquad\qquad\qquad\qquad\qquad\qquad\qquad\qquad\qquad\qquad\qquad$ (2)

$E$ signifies the number of all the electrons in the $PSS_{136}$ brush per polymer molecule, $V$ its volume. $E_{SS}$, $E_{Cs}$ and $E_{H_2O}$ correspond to the electrons of a PSS monomer (92, taking into account 90 % sulfonization), a $Cs^+$ ion (54) and a water molecule (10), respectively. The volume symbols $V_{SS}$ (200Å$^3$), $V_{Cs}$ (38Å$^3$) and $V_{H_2O}$ (30Å$^3$) are assigned in a similar way. ($\rho_i$ and $l_i$ symbolize the electron density and the thickness of the slab $i$.) $V_{Cs}$ is inferred from the density changes of aqueous solutions containing Cs salts[30]. The polymerization index is denoted by $n_{ss}$. For each molecular area $A$, $n_{H_2O}$ and $n_{Cs}$ are calculated from the above two equations. $n_{Cs}$ is found to be independent from $A$, $n_{Cs} = 67 \pm 4$ (for $n_{SS} = 136$ and calculated from 10 different measurements). Similar values for $n_{Cs}$ are found, when the subphase contains 0.1 Mol/L salt. Specifically, $n_{Cs}$=76 for 0.1 Mol/L CsCl, and $n_K$=82 for 0.1 Mol/L KCl); this means that the charge of only about every second monomer is compensated by a $Cs^+$ ion. Comparing this with the concentration ratio of $Cs^+$ ions and protons in the bulk, it follows that the presence of Caesium ions in the brush is energetically disfavored compared to protons, or in other words, that protons are sucked into the brush (which will be discussed further at the end of this section). On compression, $n_{H_2O}$ is reduced by a factor of three, from 12000 to 3750.

To calculate the $PSS_{136}$ brush height $L$ in a way which is independent of model parameterization, we followed the approach suggested by numerical simulations,[21,25] and calculated it from the first moment of $\phi_{PSS}$, the $PSS_{136}$ volume fraction:

$$L = 2 \frac{\int_0^\infty z \phi_{PSS}(z) dz}{\int_0^\infty \phi_{PSS}(z) dz} = 2 \frac{\int_0^\infty z (\rho(z) - \rho_{sub}) dz}{\int_0^\infty (\rho(z) - \rho_{sub}) dz}$$ 

$\qquad\qquad\qquad\qquad\qquad\qquad\qquad\qquad\qquad\qquad\qquad\qquad\qquad$ (3)

with the boundary $z = 0$ set at the interface between the flatly adsorbed PSS layer and the osmotically swollen PSS brush. The factor 2 in equation (3) accounts for the fact that (for a uniform density inside the brush) the brush height is twice the first moment. For

$$\phi_{PSS}(z) = \frac{\rho(z) - \rho_{sub}}{\frac{n_{SS} E_{SS} + n_{Cs} E_{Cs}}{n_{SS} V_{SS} + n_{Cs} V_{Cs}} - \rho_{sub}}$$ 

$\qquad\qquad\qquad\qquad\qquad\qquad\qquad\qquad\qquad\qquad\qquad\qquad\qquad$ (4)

the additional constraint $n_{Cs} = 0.49 n_{SS}$ (for $n_{SS}$=136) is used, based on the assumption that the $Cs^+$ to SS-monomer ratio is constant within the brush. The brush length $L$ increases slightly on compression (cf. Fig.2).

Similarly, we performed reflectivity measurements on the shorter diblock copolymer, $PEE_{114}PSS_{83}$ (cf. Fig. 1). On clean water, the contrast is not good, only one maximum is observed, and consequently the brush thickness has a quite large error bar.[1,2] However, using a subphase with 1mM CsCl for contrast enhancement, two maxima are observed. Then, three slabs are necessary to describe the electron density profile of the $PSS_{83}$ brush (one for the adsorbed PSS layer, two for the brush profile). A somewhat larger $Cs^+$ incorporation is found, $n_{Cs} = 71 \pm 3$ and $n_{Cs} = 0.85 n_{SS}$. As for the longer $PSS_{136}$ brush, on monolayer compression a small but unambiguous increase of the brush length $L$ is observed (cf. Fig.2). For the sake of consistency, reflection data measured before on clean water[1,2] are reanalyzed with the approach described above (cf. Eq. (3)). However, for these systems the brush height $L$ exhibits much larger scattering (cf. Fig. 2).

If the reflectivity data would show only one interference maximum due to the polyelectrolyte brush, or if the maxima were equidistant, one slab would be sufficient to describe the data, since we apply the maximum entropy approach.[29] This would lead to a segmental density profile with very few features. However, the pronounced structure of the data allows very detailed information to be obtained. This, in turn, necessitates a discussion, how to measure the brush thickness. The non-equidistant maxima are due to the slowly decaying segmental density towards the water, which can be modeled by a sequence of adjacent slabs, together with the corresponding interfacial roughness parameters. The fitting parameters of those adjacent PSS slabs are strongly correlated, and different combinations of parameters lead to the same density profile.[26,31,32] If we would calculate the brush thickness as the sum of the PSS slabs ($\sum_i l_{PSS,i}$)[1,2], as we and others did with less structured data, we would obtain a larger thickness and substantially more scatter. In contrast, the $1^{st}$ moment given in Eq. (3) measures the average length of the whole PSS brush, independent of the parameterization. The remarkable low scatter of this approach allows to observe unambiguously even small changes of the brush thickness, provided the data are structured.

For $PEE_{144}PSS_{136}$, the polyelectrolyte brush length $L_{N=136}$ is found to be 40 to 50% of the contour length, (340 Å). The shorter $PSS_{83}$ brush is more stretched, for the grafting density $\rho_a = 10^{-3}$Å$^{-2}$ one obtains only $L_{N=136}/L_{N=83} = 1.35$, whereas 136/83 = 1.64. If the PSS-brushes anchored to the water surface are even longer ($N$=356), the stretching amounts only to 30-40% of the contour length as shown in [11]. The brush length $L$ is fitted to Eq. (12) with the following preselected parameters: monomer length $a$=2.5 Å and polyelectrolyte block degree of polymerization $N$=136 ($N$=83, respectively). Considering Eq. (12), the decreased stretching with polymer length has to be attributed to a decrease in $f$, the effective charge fraction. Since we found from the analysis of the X-ray reflectivity data that the counterion incorporation decreases with $N$, too, we assumed that the degree of chain dissociation $f$ is identical to the degree of Cs incorporation, that is $f$=0.49 ($f$=0.85, respectively). This amounts to assuming that protons which are incorporated into the layer are either chemically or electrostatically bound to the charged monomers. Then, the only free parameter in the fits to Eq. (12) is $\sigma_{eff}$.

Very similar values are obtained for $\sigma_{eff}$ for both PSS lengths: 13.84 Å and 13.96 Å respectively, leading to a minimum chain area of $\sigma_{eff}^2$ =193Å$^2$. This value is of the same order as the minimum area which a compressed PSS chain with a radius of $r$=6Å occupies, $i.e.$ $\pi r^2$/0.91=124Å$^2$ (assuming a hexagonal lattice). Clearly, the fitted effective excluded volume is larger since it also takes into account the finite volume of the counterions which might or might not be hydrated.

Note that both values obtained for the effective chain dissociation ($f$=0.85 for $N$=83 and $f$=0.49 for $N$=136) exceed $f$=0.33, a value which one would obtain assuming the classical Manning condensation of counterions,[33] and which we did indeed obtain, when we fitted the length of the $PSS_{356}$ brushes described in [11]. It was shown theoretically by using the full Poisson-Boltzmann equation that non-linear effects (associated with Manning condensation) indeed become less pronounced in highly condensed polyelectrolyte brushes.[18] A detailed discussion of the degree of chain dissociation exceeds the scope of the present paper, yet some conclusions obtained from the experiments are apparent: (ı) the degree of chain dissociation exceeds for short polymers the one predicted by the classical Manning condensation and (ıı) the freely moving counterions within an osmotically swollen brush are those which occur in excess within the subphase. It is important to note that the fitted chain dissociation constant $f$ reflects both electrostatic counterion-condensation effects (equivalent to Manning condensation) and chemical charge-regulation effects (which are described by a local association-dissociation equilibrium of the acidic groups). The kind and concentration of the bound counterions are determined by a subtle balance of the chemical potential within and outside the brush, and the binding constants between the polyelectrolyte and the respective counterions. Obviously, in the case of PSS, the proton binding constant exceeds the Cs-binding constant substantially. A further complication arises due to the ion-specific hydration structure of ions which will be also modified in a highly condensed environment such as the brush we are studying. All these factors make up ion-specific effects such as categorized in the famous Hofmeister series.

## 3. Simulations

Computer simulations provide an excellent mean to study polymer systems. Extensive molecular dynamics simulations have been performed recently on polyelectrolyte brushes at various grafting densities and charge fractions, both at strong and intermediate electrostatic couplings.[19-21] In these simulations, a freely-jointed bead-chain model is adopted for polymers end-grafted onto a rigid surface. The short-range repulsion between particles separated by distance $r$ is modeled by a shifted Lennard-Jones (LJ) potential $u_{LJ}(r) = 4\varepsilon_{LJ}\left\{(\sigma/r)^{12} - (\sigma/r)^6 + 1/4\right\}$ for $r/\sigma < 2^{1/6}$ and $u(r) = 0$ otherwise, with Lennard-Jones diameter of $\sigma$ being equal for both counterions and monomers $\sigma = \sigma_m = \sigma_{ci}$. The monomers are connected by non-linear springs with the so-called FENE (finite extensible non-linear elastic) potential $u_{FENE}(r) = -kR_0^2/2 \log\left\{1 - (r/R_0)^2\right\}$ for $r < R_0$ and $u_{FENE}(r) = 0$ otherwise, where the bond strength is $k = 30\varepsilon_{LJ}/\sigma^2$ and the maximum bond length $R_0 = 1.5\sigma$. For completely charged chains, this choice of parameters gives an average bond length $a$=0.98σ. The counterions are explicitly modeled as charged particles where both counterions and charged monomers are univalent and interact with the bare Coulomb potential $u_{Coul}(r) = k_B T q_i q_j l_B / r$, with $q_i, q_j = \pm 1$ being the two charges. The strength of the Coulomb interaction is given by the Bjerrum length $\ell_B = e^2/(4\pi\varepsilon_0\varepsilon k_B T)$ which measures the distance at which two elementary charges interact with the thermal energy $k_B T$; in water at room temperature, one has $\ell_B \approx 0.7 nm$. No additional electrolyte is added. The simulation box is periodic in lateral directions and finite in the $z$-direction normal to the anchoring surface at $z=0$. We apply the MMM technique introduced by Strebel and Sperb[34] and modified by Arnold and Holm[35] for laterally periodic systems (MMM2D) to account for the long-range nature of the Coulomb interactions. To study the system in equilibrium we use stochastic molecular simulation at constant temperature.

Figure 3 shows a snapshot from the simulation at an electrostatically intermediate coupling strength ($\ell_B = \sigma$) of a brush with 36 chains of 30 monomers, which is fully charged and has a large grafting density of $\rho_a = 0.12\sigma^{-2}$. Simulated density profiles of monomers and counterions of the system in normal direction are shown in Figure 4 for the fully charged brush at several grafting densities. As seen, both monomers and counterions follow very similar nearly-step-like profiles with uniform amplitude inside the brush, which increases with grafting density. These figures show that the counterions are mostly confined in the brush layer and that the electroneutrality condition is satisfied locally. A simple explanation of this finding is presented in the scaling section. One may observe (more clearly from the mean brush heights extracted from simulations and shown in Figure 5) that the polyelectrolyte chains are stretched up to about 60% of their contour length, and thus their elastic behavior is far beyond the linear regime. Therefore, within the chosen range of parameters, the simulated brush is in the strong-charging and strong-stretching limits and as we will show below, it exhibits the non-linear osmotic brush regime. The average height of end-points of the chains is one of the quantities that can be directly measured in the simulations and is shown in Figure 5 together with the predictions of our simple scaling theory to be explained further below. It is observed that the simulated brush height (solid circles) varies slowly with the grafting density, contrary to the predictions of standard scaling theories[12,13] but in agreement with the experimental results and in agreement with the scaling theory that incorporates non-linear elastic and osmotic effects. Note that the simulation model has not been adapted to model a certain polymer, but to cover generic features. In particular, the volume interaction is included by a purely repulsive Lennard-Jones potential which models the behavior in a good solvent. From theory one knows, that the particular stretching behavior of the brush depends on the relation between second virial and Bjerrum length.[20] Therefore a more refined parameter mapping is desired before discussing any specific polymer quantitatively.

## 4. Scaling Theory

An analytical theory for polyelectrolyte brushes relies on a number of simplifying assumptions. The full theoretical problem is intractable because the degrees of freedom of the polymer chains and the counterions are coupled. It is important to note that charged polymers by itself are not fully understood, therefore quite drastic simplification are needed to tackle the more complicated system of polyelectrolytes end-grafted to a surface[36].

Firstly, we will exclusively consider polymeric systems with counterions, i.e. we will neglect the presence of additional added salt. Secondly, we will write the total free energy per unit area

$$F = F_{chain} + F_{ion} + F_{int} \tag{5}$$

as a sum of separate contributions from the polyelectrolytes, $F_{chain}$, contributions from the counterions, $F_{ion}$, and an electrostatic interaction term which couples polymers and counterions, $F_{int}$. The schematic geometry of the brush system is visualized in Figure 6: the polymer chains are assumed to extend to a distance L from the grafting surface, the counterions in general form a layer with a thickness of H. As we will show in the following, the counterion layer height is typically very close to the polymer layer height.

The main contribution to the polymer free energy comes from the elastic response due to the stretching of chains. For a freely jointed chain, the entropy loss due to stretching can be calculated exactly [18]. We only need here the asymptotic expressions for weak and for strong stretching, which read (given per unit area)

$$F_{chain}/k_B T \equiv \begin{cases} 3\rho_a L^2/(2Na^2) & for \quad L << Na \\ -N\rho_a \ln(1 - L/Na) & for \quad L \approx Na \end{cases} \tag{6}$$

which are proportional to the grafting density $\rho_a$. $N$ denotes the polymerization index of the chain and $a$ is the effective Kuhn length. The contour length of a fully stretched chain follows as $Na$. The weak-stretching term is the standard term used in previous scaling models[12,13]. For the highly stretched situations encountered in fully charged brushes, the strong-stretching term valid for $L \approx Na$ is typically more appropriate and leads to a few changes in the results as will be explained further below.

The counterion free energy contains entropic contributions (due to the confinement of the counterions inside a layer of thickness H) and also energetic contributions which come from interactions between counterions. In previous theories, a low-density expansion for the interaction part was used. Most notably, the second-virial interaction is important for neutral brushes and is the driving repulsive force balancing the chain elasticity.[12,13,16,20] For charged brushes, the leading term of the electrostatic correlation energy (which shows fractional scaling with respect to the charge density) is attractive and has been shown to lead to a collapse of the polyelectrolyte brush for large Bjerrum lengths.[20] In the present analysis we use a free-volume approximation very much in the spirit of the van-der-Waals equation for the liquid-gas transition. For this we concentrate on the effective hard-core volume of a single polyelectrolyte chain, which we call $v$, and which reduces the free volume that is available for the counterions. This free volume theory therefore takes the hard-core interactions between the polymer monomers and the counterions into account in a non-linear fashion. Compared to that, the excluded-volume interaction between counterions is small since the monomers are more bulky than the counterions and therefore it is neglected.[18] The non-linear entropic free energy contribution of the counterions reads

$$F_{ion} / k_B T = \rho_a N f \ \ln \frac{\rho_a N f}{H - \rho_a v} \ - 1 \tag{7}$$

where $f$ is the charge fraction of the chains. In the limit of vanishing polymer self-volume, $v \to 0$, one recovers the standard ideal entropy expression. As the volume available for the counterions in the brush, which per polymer is just $H / \rho_a$, approaches the self volume of the polymers, $v$, the free energy expression (7) diverges, that means, the entropic prize for that scenario becomes infinitely large. The self volume of the polymers is roughly independent of the polymer brush height, and can be written in terms of the monomer hard-core diameter $\sigma_{eff}$ and the polymer contour length $aN$ as, $v = aN\sigma_{eff}^2$, where $\sigma_{eff}$ is the sum of the monomer and counterion diameters, $i.e.$ $\sigma_{eff} = \sigma_m + \sigma_{ci}$. This leads to the final expression [18]

$$F_{ion} / k_B T = \rho_a N f \left[ \ln \left( \frac{\rho_a N f}{H - \rho_a \sigma_{eff}^2 Na} \right) - 1 \right] \tag{8}$$

Finally, the electrostatic interaction between polyelectrolytes and counterions is considered on the mean-field level, where the charges are smeared out over the brush region (0<z<L) and over the polymer-free counterion region (L<z<H). It reads [20]

$$F_{int} / k_B T = (2\pi / 3) \ell_B (\rho_a N f)^2 (H - L)^2 / H \tag{9}$$

An electrostatic coupling only arises when the counterion layer extends over the polymer-brush layer, i.e. when $H > L$, and it is the driving force that keeps the counterions inside the brush layer. This force will be important later in order to estimate when counterions will start to leave the brush.

*The standard osmotic brush regime* results from balancing the elastic stretching term for small stretching, L<<Na, Equation (6), with the counterion entropy in the absence of a polymer self volume, which is Equation (8) in the limit $\sigma_{eff} = 0$ and for H=L. The result from minimizing the resulting free energy with respect to the brush height $L$ is the classical result[12,13]

$$L \cong aN\sqrt{f/3} \qquad (10)$$

The main assumption here is that the counterions all stay inside the brush, that is H=L, which will be critically examined at the end of this section. Also, it is clear that for highly charged polymers, i.e. for f=1, the predicted stretching in Equation (10) goes beyond the assumption of weak stretching.

In *the strongly-stretched osmotic brush regime*, one chooses the strong-stretching version of the chain-stretching entropy in Equation (6) and balances it with the counterion entropy for vanishing polymer self volume for the case *H=L.* The result is[18]

$$L \cong aNf/(1+f) \qquad (11)$$

which is the large-stretching analogue of Equation (10). The maximal stretching predicted from this equation is obtained for *f=1* and corresponds to a stretching of 50% of the contour length. This height is considerably smaller than what is observed in simulations and experiments. Moreover, the predicted brush height in Equation (11) does *not* depend on the grafting density. (For comparison, both expressions (10) and (11) are shown in Figure 5 as dashed lines (a) and (b) respectively.) It transpires that something is missing in the above scaling description. This something, we propose, is the entropic pressure which increases as the volume within the brush is progressively more filled up by the polymer self–volume (see below). The fact that the non-linear elastic stretching of the chains by itself does not lead to a grafting-density dependence for the brush height has also been noted in previous studies.[22,23]

In the *non-linear osmotic brush regime* we combine the high-stretching (non-linear) version of the chain elasticity in Equation (6) with the non-linear entropic effects of the counterions due to the finite volume of the polymer chains, i.e. we choose a finite polymer diameter in Equation (8). The final result for the equilibrium brush height is[18]

$$L \cong aN(f + \sigma_{eff}^2 \rho_a)/(1+f) \qquad (12)$$

which in the limit of maximal grafting density, that is close packing, $\rho_a = 1/\sigma_{eff}^2$, reaches the maximal value $L = Na$, as one would expect: Compressing the brush laterally increases the vertical height and finally leads to a totally extended chain structure. In Figure 5, we compare expression (12), shown as a solid line, with simulation results for the brush height as a function of grafting density. Note that we have used $\sigma_{eff}^2 = 2\sigma^2$, where $\sigma = \sigma_m = \sigma_{ci}$ is the LJ diameter of monomers and counterions in the simulations. This choice corresponds to an approximate two-dimensional square-lattice packing of monomers and counterions on two interpenetrating sublattices. The scaling prediction, Eq. (12), qualitatively captures the slow increase of the brush height with grafting density. The deviations from the simulation data may be explained by considering additional effects, such as lateral inhomogeneity of counterion distribution around the brush chains and intermediate-stretching elasticity of the chains. They are not included in Eq. (6) and go beyond the present scaling analysis, as will be discussed briefly at the end of this section.[18] Also, the simulation is not dealing with hard spheres but with soft potentials, which will modify the ratio $\sigma_{eff}/\sigma$. Note that in the comparison made with the experiments in Figure 2, we have used Eq. (12) with $\sigma_{eff}$ serving as a fit parameter to account for the effective area occupied by a compressed polymer chain.

One main approximation in the theoretical analysis is the assumption that all counterions stay localized in the brush layer. This will be analyzed critically in the following. To get a feeling for the involved forces, we will first consider the confinement of a layer of counterions at a planar charged surface. We balance the electrostatic interaction energy Equation (9) for an infinitely thin brush layer L=0 with the confinement entropy Equation (8) for $\sigma_{eff}=0$ and obtain $H = 3/(2\pi\ell_B N f \rho_a) = 3\lambda_{GC}$, which has the same scaling as the Gouy-Chapman length $\lambda_{GC} = 1/(2\pi\ell_B N f \rho_a)$, a measure of the extend of counterion layers. This shows that the scaling approach reproduces the result from the exact analysis of the Poisson-Boltzmann approach. Now we ask what the counterion-layer height is in the case of a finite brush height L. We therefore minimize the sum of the electrostatic interaction energy Equation (9) and the counterion confinement entropy Equation (8) with respect to the brush height L for vanishing polymer diameter $\sigma_{eff}=0$ and obtain the result (to first order in powers of [H-L]/L)

$$H = L + 3\lambda_{GC}/2 \qquad (13)$$

This gives the counterion layer height corresponding to the results in Equations (10) and (11) for the brush height, that is, in the absence of a finite polymer self-volume. Since for a typical fully charged brush the Gouy-Chapman length $\lambda_{GC}$ is of the order of one Angstrom or less, the counterion layer basically has the same height as the brush layer. In other words, the counterions are completely trapped inside the brush for vanishing polymer radius, in agreement with the simulation results shown in Figure 4. Now we have to do the same estimate for finite polymer radius. We minimize the sum of the electrostatic interaction energy Equation (9) and the counterion confinement entropy Equation (8) for finite L and finite polymer size $\sigma_{eff}$ and obtain the result (to first order in powers of [H-L]/L)

$$H = L + \frac{3\lambda_{GC}}{2(1-\eta)} \qquad (14)$$

where $\eta = \rho_a \sigma_{eff}^2 N a / L$ measures the ratio of the polymer excluded volume $v = \sigma_{eff}^2 N a$ and the volume in the brush available for a polymer $L/\rho_a$ and thus the degree of close-packing in the brush. For a grafting density of $\rho_a = 0.1 nm^{-2}$, a polymer length of $L=15 nm$ and a monomer number of $N=136$, as used in the experiments, one obtains a Gouy-Chapman length of $\lambda_{GC} = 1/(2\pi\ell_B N f \rho_a) \approx 0.01 nm$. Therefore, even for a close packing fraction of 99%, i.e. for $\eta = 0.99$, the difference between the counterion layer height and the brush height is only about a nanometer, which is rather negligible compared to the total brush height. This argument reflects the strong electrostatic interaction between the brush and the counterion layer, and it shows that the underlying assumption of local electroneutrality in Equation (12) is justified, even for cases where the non-linear osmotic pressure is rather large and leads to a brush height very close to full extension of the chains.

Another mechanism which may play a role in the non-linear osmotic brush regime and is neglected at the present scaling level here is associated with lateral electrostatic effects due to, for instance, the lateral inhomogeneous distribution of counterions around the brush chains. Such inhomogeneities have indeed been found both in simulations[19,21] and in experiments. [1,4,5] For decreasing grafting density, such effects do become important and lead to non-monotonic behavior of the brush height as a function of grafting density. These aspects have been studied in detail in Ref. [18] using the full non-linear Poisson-Boltzmann equation. For the large grafting density regime considered here, however, lateral electrostatic effects may be neglected at the scaling level. Chemical binding effects of counterions to the monomers are not explicitly considered in our theoretical modeling. However, in the comparison made with the experimental data, we have used the effective charge fraction of the chains in the scaling prediction, Eq. (12), to account for chemical ion binding to polyelectrolyte chains (see Section 2) in a simple manner.

## 5. Conclusions

New experiments on highly compressed and highly charged polyelectrolyte brushes at the air-water interface are presented. X-ray reflectivity measurements suggest the brush height $L$ to slightly increase with increasing grafting density, in contrast to leading-order scaling results for the osmotic brush regime. This slight dependence of brush height on grafting density is reproduced in Molecular-Dynamics simulations which take the electrostatic interactions between monomers and counterions and the excluded-volume effects into account. A simple scaling picture is presented which incorporates the excluded-volume effects in a free-volume formulation, similar to the classical derivation of the van-der-Waals equation of state and which can qualitatively describe these findings.

## Figure Captions

Fig.1: (a) Normalized X-ray reflectivity measurements of the monolayer $PEE_{144}PSS_{136}$ on a 1mM CsCl aqueous solution along the isotherm (shown in the inset, $\pi_1$ indicates the formation of a homogeneous PEE layer) at different molecular areas. The lines are fits. For clarity, each reflectivity curve is displaced by 0.4. (b) Same for $PEE_{144}PSS_{83}$.

Fig.2: The thickness $L$ of the osmotically swollen $PSS_{136}$ (circles) and $PSS_{83}$ (squares) brush vs. grafting density. The subphase contained either 1mM CsCl (filled symbols) or was just clean water at pH=5.5 (open symbols). The dashed lines are power-law fits with the exponents 0.17 and 0.13, respectively. The solid lines are fits to Eq. (12) with the following preselected parameters: monomer length $a$=2.5 Å, polyelectrolyte block degree of polymerization $N$=136 ($N$=83, respectively) and the effective chain dissociation $f$ from the counterion incorporation determined by X-ray reflectivity. The degree of chain dissociation $f$=0.49 ($f$=0.85, respectively) was found to be identical to $Cs^+$-incorporation. Fits were only made to the filled symbols. The free parameter in the fits is $\sigma_{eff}$, for which very similar values are obtained: 13.84 Å and 13.96 Å, respectively.

Fig 3: A snapshot of a polyelectrolyte brush with 36 chains of N=30 monomers (in light gray) from MD simulations at electrostatically intermediate coupling strength $(\ell_B = \sigma)$. Chains are fully charged and anchored at grafting density of $\rho_a = 0.12\sigma^{-2}$. Counterions are shown by dark grey spheres. The box height perpendicular to the anchoring plane has been reduced for the sake of representation.

Fig 4: Simulated density profiles of monomers $\rho_m(z)$ (open symbols) and counterions $\rho_{ci}(z)$ (filled symbols) as a function of the distance from the anchoring surface. Shown are profiles

for fully charged brushes of 36 chains of N=30 monomers with $\ell_B = \sigma$ at grafting densities (from bottom to top) $\rho_a \sigma^2 = 0.020$ (triangles left)  0.042  (circles), 0.063 (squares), 0.094 (diamonds), and 0.12 (triangles top).

Fig 5: Brush height as a function of grafting density for polyelectrolyte chains of N=30 monomers with charge fraction f=1. Symbols show simulation data with corresponding linear fit (dot-dashed line), the straight solid line represents the prediction of the non-linear scaling theory, Eq. (12), with $\sigma_{eff}^2 = 2\sigma^2$ . The dashed lines (a) and (b) show the scaling predictions Equations (10) and (11) respectively.

Fig 6: The schematic geometry of a charged brush with $L$ and $H$ denoting the brush height and the height of the counterion layer, respectively.

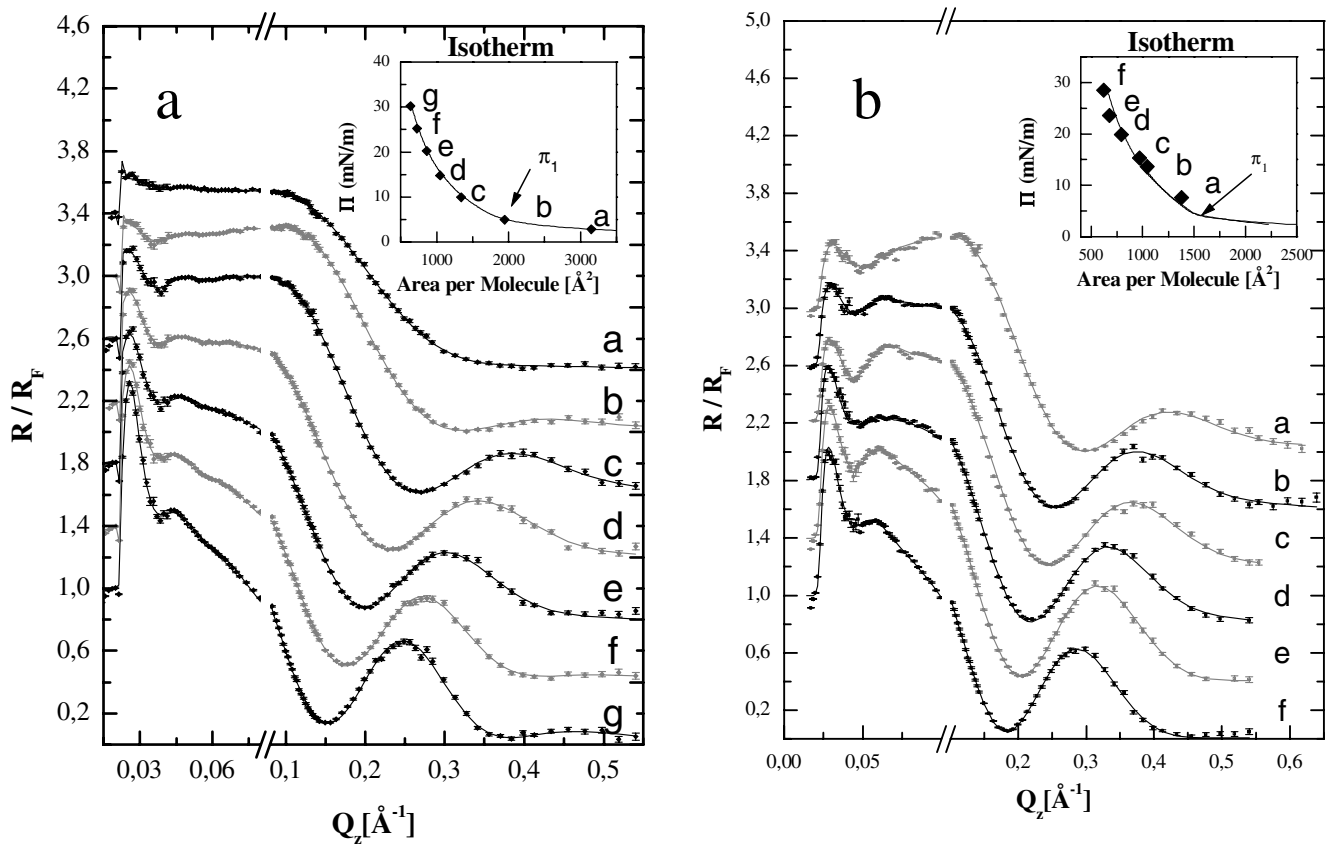

Fig.1: (a) Normalized X-ray reflectivity measurements of the monolayer PEE$_{144}$PSS$_{136}$ on a 1mM CsCl aqueous solution along the isotherm (shown in the inset, $\pi_1$ indicates the formation of a homogeneous PEE layer) at different molecular areas. The lines are fits. For clarity, each reflectivity curve is displaced by 0.4. (b) Same for PEE$_{144}$PSS$_{83}$.

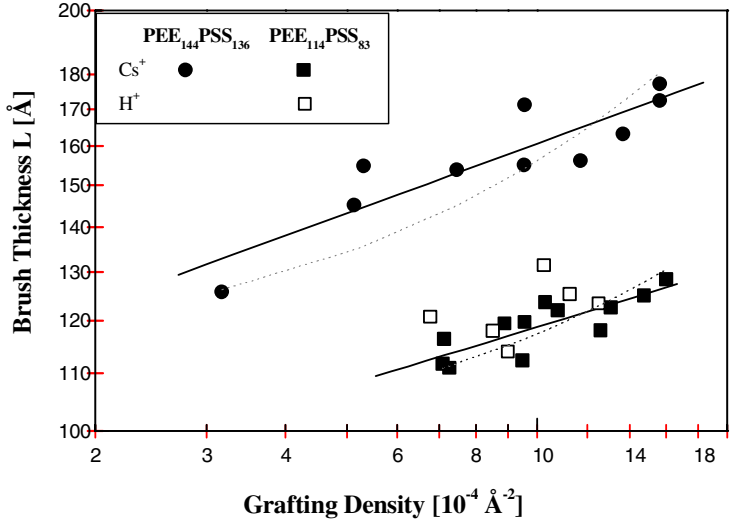

Fig.2: The thickness $L$ of the osmotically swollen $PSS_{136}$ (circles) and $PSS_{83}$ (squares) brush vs. grafting density. The subphase contained either 1mM CsCl (filled symbols) or was just clean water at pH=5.5 (open symbols). The dashed lines are power-law fits with the exponents 0.17 and 0.13, respectively. The solid lines are fits to Eq. (12) with the following preselected parameters: monomer length $a$=2.5 Å, polyelectrolyte block degree of polymerization $N$=136 ($N$=83, respectively) and the effective chain dissociation $f$ from the counterion incorporation determined by X-ray reflectivity. The degree of chain dissociation $f$=0.49 ($f$=0.85, respectively) was found to be identical to $Cs^+$-incorporation. Fits were only made to the filled symbols. The free parameter in the fits is $\sigma_{eff}$, for which very similar values are obtained: 13.84 Å and 13.96 Å, respectively.

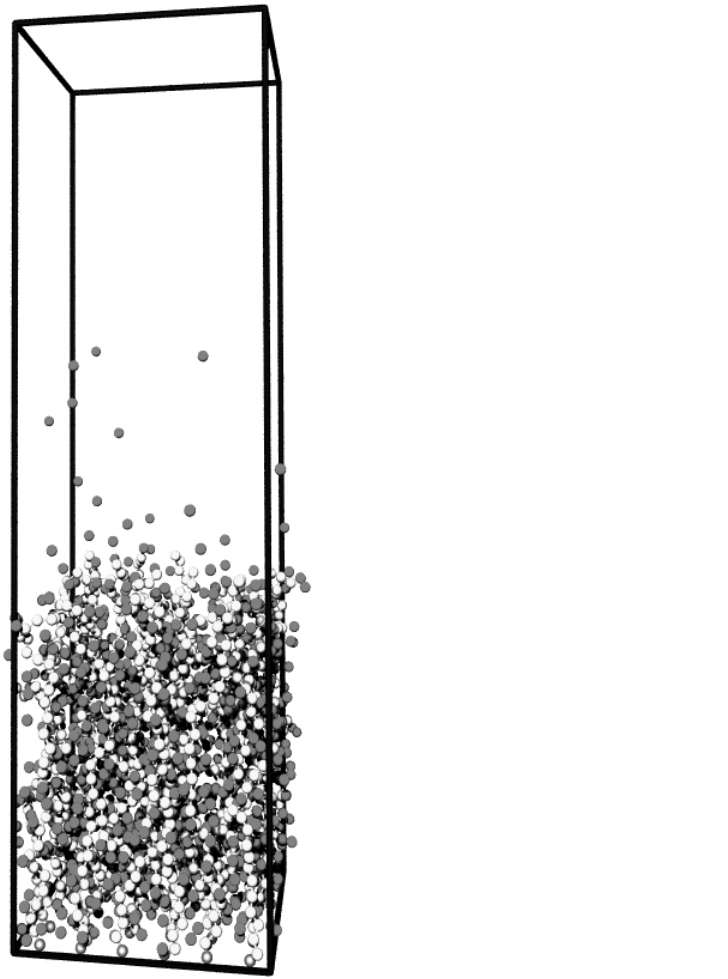

Fig 3: A snapshot of a polyelectrolyte brush with 36 chains of N=30 monomers (in light gray) from MD simulations at electrostatically intermediate coupling strength $(\ell_B = \sigma)$. Chains are fully charged and anchored at grafting density of $\rho_a = 0.12\sigma^{-2}$. Counterions are shown by dark grey spheres. The box height perpendicular to the anchoring plane has been reduced for the sake of representation.
The Figure is shown twice, once as a tif-file, and once as a jpg.file. Please choose the one which is more convenient.

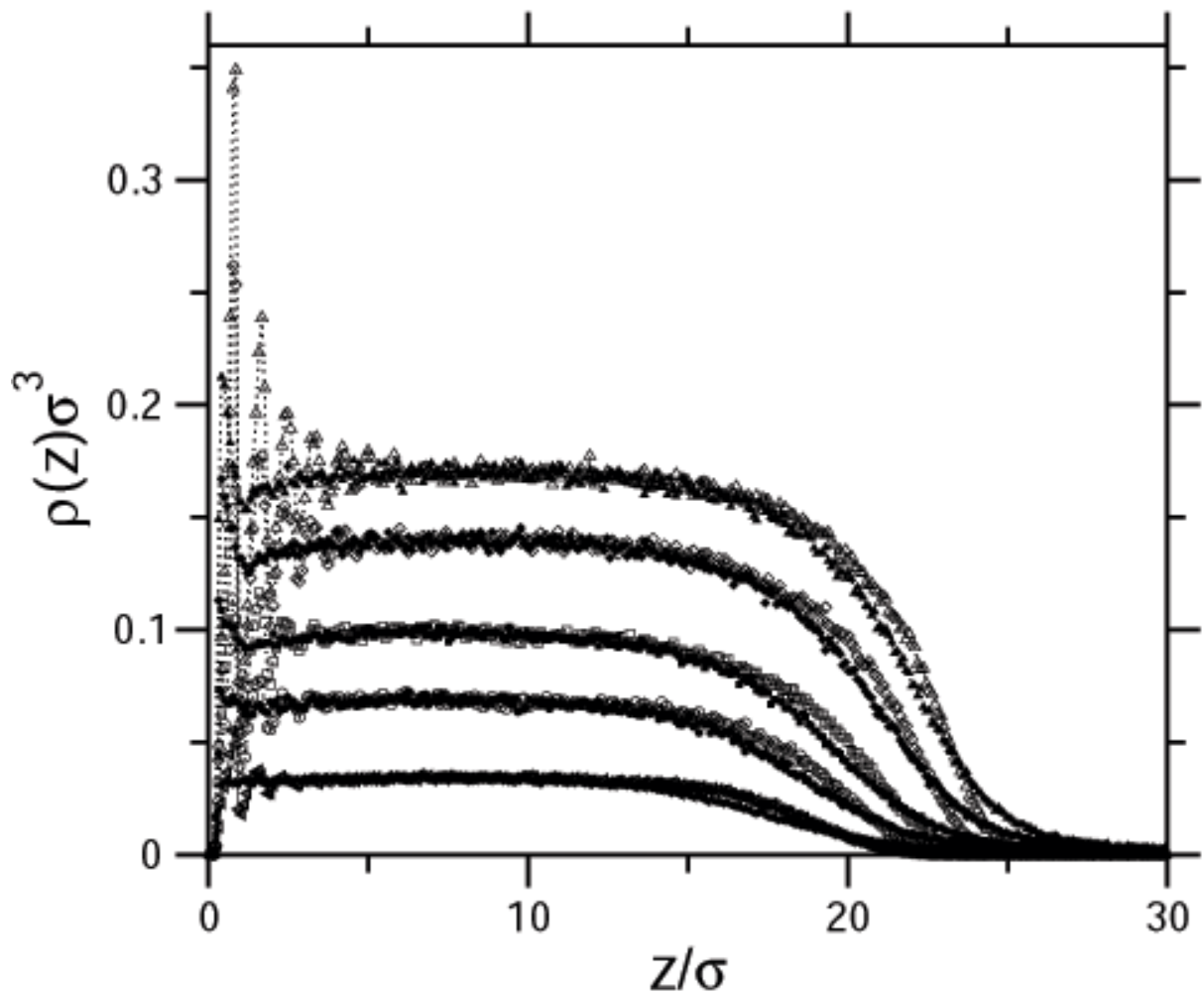

Fig 4: Simulated density profiles of monomers $\rho_m(z)$ (open symbols) and counterions $\rho_{ci}(z)$ (filled symbols) as a function of the distance from the anchoring surface. Shown are profiles for fully charged brushes of 36 chains of N=30 monomers with $\ell_B = c$ at grafting densities (from bottom to top) $\rho_a\sigma^2 = 0.020$ (triangles left) 0.042 (circles), 0.063 (squares), 0.094 (diamonds), and 0.12 (triangles top).

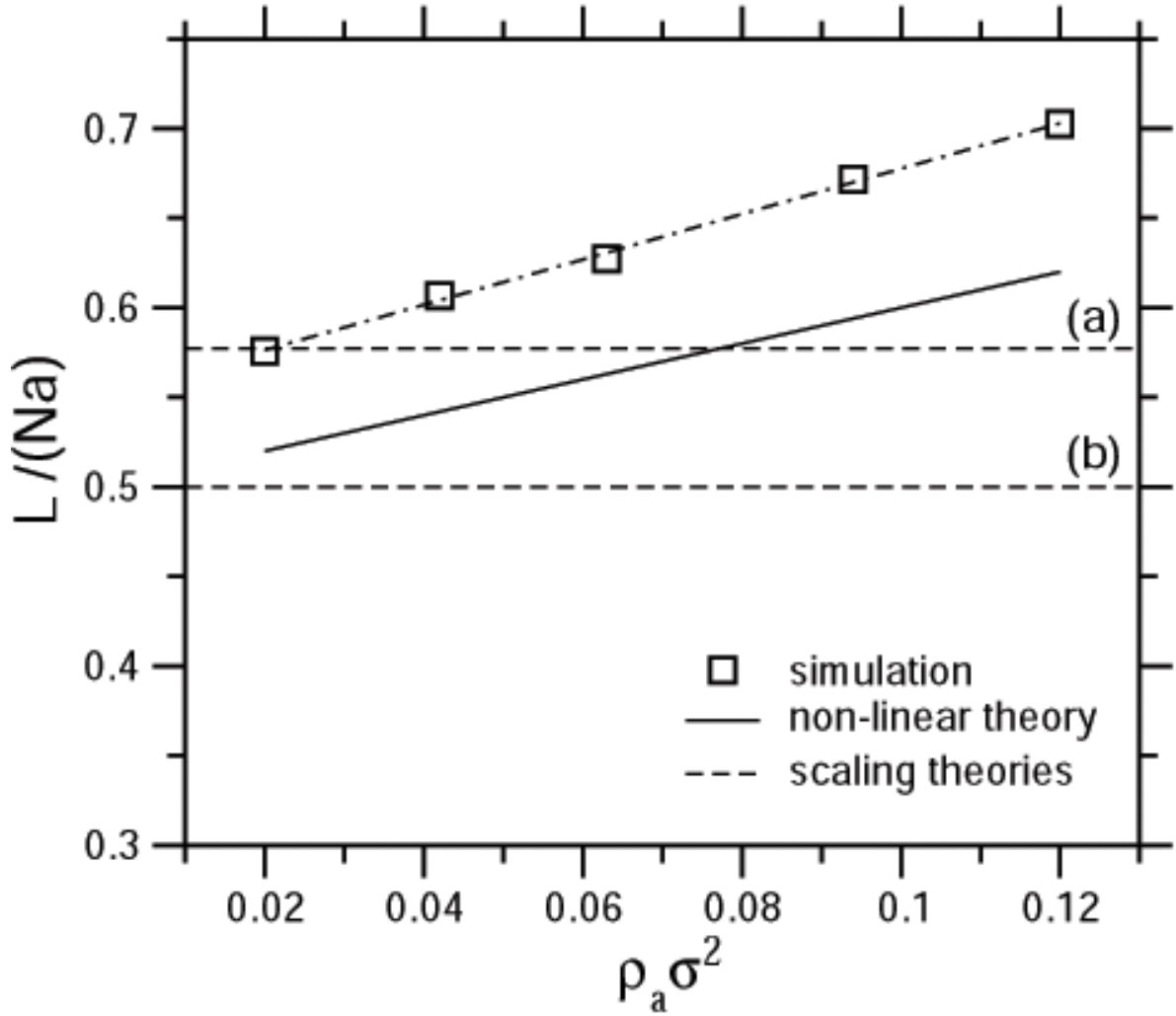

Fig 5: Brush height as a function of grafting density for polyelectrolyte chains of N=30 monomers with charge fraction f=1. Symbols show simulation data with corresponding linear fit (dot-dashed line), the straight solid line represents the prediction of the non-linear scaling theory, Eq. (12), with $\sigma_{eff}^2 = 2\sigma^2$. The dashed lines (a) and (b) show the scaling predictions Equations (10) and (11) respectively.

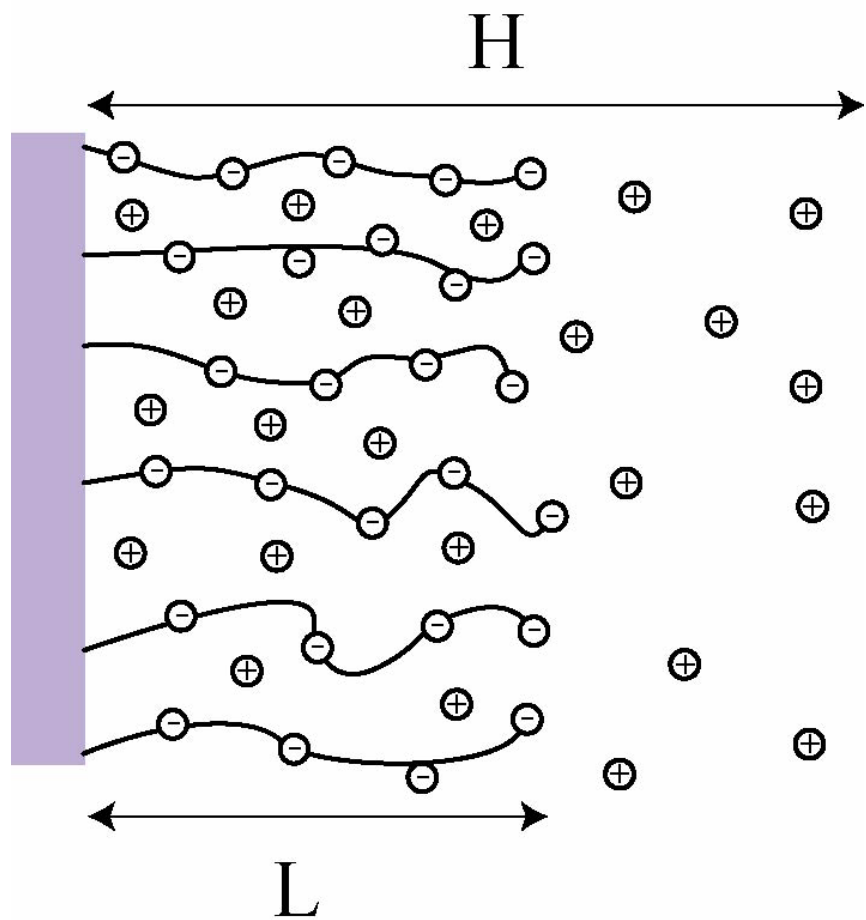

Fig 6: The schematic geometry of a charged brush with *L* and *H* denoting the brush height and the height of the counterion layer, respectively.z